\documentstyle[12pt]{article}
\newcommand{\mysection}{\setcounter{equation}{0}\section}

\begin{document}
\begin {flushright}
ITP-SB-96-7
\end {flushright} 
\vspace{3mm}
\begin{center}
{\Large \bf On the Resummation of Singular Distributions 
in QCD Hard Scattering} 
\end{center}
\vspace{2mm}
\begin{center}
Nikolaos Kidonakis and George Sterman  \\
\vspace{2mm}
{\it Institute for Theoretical Physics\\
State University of New York at Stony Brook\\
Stony Brook, NY 11794-3840, USA} \\  
\vspace{2mm}
April 1996
\end{center}

\begin{abstract}
We discuss the resummation of distributions that are singular
at the elastic limit of partonic phase space
(partonic threshold) in
QCD hard-scattering cross sections, such as  heavy quark production.
We show how nonleading soft logarithms exponentiate 
in a manner that depends on the
color structure within the underlying hard scattering.  This result
generalizes the resummation of threshold singularities for the
Drell-Yan process, in which the hard scattering proceeds through
color-singlet annihilation.  We illustrate our results for the
case of heavy quark production by light quark annihilation,
and briefly discuss its extension to heavy quark
production through gluon fusion.
\end{abstract}

\pagebreak

\mysection{General Formalism}

In hard scattering cross sections factorized according to perturbative 
QCD the calculable short-distance function includes distributions that 
are singular when the total invariant mass of the partons reaches the
minimal value necessary to produce the observed final state. Such singular
distributions can give substantial QCD corrections to any order in $\alpha_s$. 

Expressions that resum these distributions in the short-distance 
functions of Drell-Yan cross sections to arbitrary logarithmic accuracy
have been known for some time \cite{oldDY}.   
It has also been observed that leading distributions,
and hence leading logarithms in moment space, are the same for
many hard QCD cross sections.  This has been used as the
basis for important estimates of heavy quark 
production including resummed
leading, and some nonleading,
logarithms \cite{heavyquarkresum}.  
In this paper, we shall exhibit a 
 method by which nonleading distributions may be treated, and illustrate
this method in the case of heavy quark production through the
annihilation of light quarks.

We consider the inclusive cross section for the production of 
one or more particles, with total invariant
mass $Q$.  Examples include states produced by
QCD, such as heavy quark pairs \cite{heavycalcs} or
high-$p_T$ jets \cite{jetcalcs}, in addition to massive 
electroweak vector bosons, virtual or real, as in the Drell-Yan
process.  

To be specific,  
we shall discuss the summation of (``plus")
distributions, which are singular  for $z=1$, where
\begin{equation}
z={Q^2\over s},
\label{def}
\end{equation}
for the production of a heavy quark pair of total invariant
mass $Q$,
with $s$ the invariant mass squared
of the incoming partons that initiate the hard scattering.  
We shall refer to $z=1$ as  ``partonic threshold" 
\footnote{We emphasize that by partonic threshold, we
refer to c.m.\ total energy of the incoming partons
for a fixed final state; heavy
quarks, for example, are not necessarily produced at rest.},
or more accurately the ``elastic limit.''  
We assume that the cross section is defined so that
there are no uncancelled collinear divergences in the final state.  

The main complications relative to Drell-Yan \cite{oldDY} involve
the exchange of color in the hard scattering, and the presence of
final-state interactions.  In fact, these effects only
modify partonic threshold singularities at next-to-leading logarithm,  and we
give below explicit exponentiated
moment-space expressions which take them into account
at this level.  At the next level of accuracy, we shall see that resummation
requires ordered exponentials, in terms of calculable anomalous dimensions.

The properties of QCD that make this organization possible are the factorization
of soft gluons from high-energy partons in perturbation theory \cite{fact}, and the
exponentiation of soft gluon effects \cite{expon}.  
Factorization is represented by
Fig.\ 1 for the annihilation of a light quark pair to form a pair of
heavy quarks.  In this figure, momentum configurations that contribute singular
behavior near partonic threshold
are shown in a cut diagram notation \cite{fact}.
As shown, it is possible to factorize soft gluons from
the ``jets" of virtual and real particles that are on-shell and
parallel to the 
incoming, energetic light quarks, as well as from the outgoing heavy
quarks. Soft-gluon factorization 
from incoming light-like partons is 
a result of the ultrarelativistic limit \cite{fact},
while factorization from heavy quarks, even when they
are nonrelativistic, is familiar from heavy-quark effective
theory.  Once soft gluons are factored from them, the jets
may be identified with parton distributions of the
initial state hadrons.  The
hard interactions, labelled $H_I$ and $H^*_J$ in the figure, corresponding
to contributions from the amplitude and its complex conjugate respectively,
are labelled by the overall color exchange in each.
A general argument of how the exponentiation of Sudakov logarithms follows
from the factorization of soft and hard parts and jets is given in 
Ref.\ \cite{CLS}.

For example,
with the quark-antiquark process shown, the choice of color structure is simple,
and may, for instance, be chosen as singlet or octet.  To make these choices
explicit, we label the colors of the incoming pair $i$ and $j$ for
the quark and antiquark respectively, and of
the outgoing (massive) pair $k$ and $l$ for the quark and antiquark.  
The hard scattering is then of the generic form
\begin{equation}
H_1=h_1(Q^2/\mu^2,\alpha_s(\mu^2))\, \delta_{ji}\delta_{lk}\, ,
\label{singletdef}
\end{equation}
for singlet structure (annihilation of color) in the $s$-channel.
For the $s$-channel octet, or more generally adjoint in color SU(N), 
we have, analogously
\begin{equation}
H_A=h_A(Q^2/\mu^2,\alpha_s(\mu^2))\, 
\sum_{c=1}^{N^2-1}\left [ T_c^{(F)}\right]_{ji}\; 
\left [ T_c^{(F)}\right]_{lk}\, ,
\label{octetdef}
\end{equation}
with $T_c^{(F)}$ the  generators in the fundamental representation.
The functions $h_I$ are, as indicated, infrared safe, that is, free of 
both collinear and infrared divergences, even at partonic threshold.  

Taking into account possible choices of $H_I$ and $H^*_J$, 
an expression that organizes all singular distributions
for heavy quark production is
\begin{eqnarray}
\frac{d\sigma_{h_1  h_2}}{dQ^2 \; d\cos\theta^*\; dy}&=&\sum_{ab} \; \sum_{IJ}
\int\frac{dx_a}{x_a}\frac{dx_b}{x_b} \phi_{a/h_1}(x_a,Q^2) 
\phi_{b/h_2}(x_b,Q^2)
\nonumber \\ &&
\times \delta\left(y-\frac{1}{2}\ln\frac{x_a}{x_b}\right) 
\ \Omega_{ab}^{(IJ)}\left(\frac{Q^2}{x_a x_b S},y,\theta^*,\alpha_s(Q^2)\right)\, ,
\nonumber\\
\label{Qyfact}
\end{eqnarray}
where $y$ is the pair rapidity and $\theta^*$ the scattering angle
in the pair center of mass frame.
The indices $I$ and $J$ label color tensors,
such as the singlet (\ref{singletdef}) and octet (\ref{octetdef}), into which
we contract
the colors of the incoming and outgoing partons
that participate in the hard scattering. 
The variable $S$ is the invariant mass squared of the incoming hadrons. 
The functions $\phi_{a/h}$ are parton densities,
evaluated at scale $Q^2$.  
The function $\Omega$ contains all singular behavior
in the threshold limit, $z\rightarrow 1$.

$\Omega$ depends on the scheme in which we perform factorization,
the usual choices being $\overline{\rm MS}$ and DIS.  
Note that the
resummation may be carried out at fixed 
rapidity $y$, so long as $y$ is not close to
the edge of phase space \cite{LaenenSterman}. 

The color structure of the hard scattering influences contributions to
nonleading infrared behavior.  Not all soft gluons, however, are sensitive
to the color structure of the hard scattering.  Gluons that are both soft
and collinear to the incoming partons
factorize into the parton distributions of the incoming hadrons. It is at
the level of nonleading, purely soft gluons with central rapidities that 
color dependence appears
in the resummation of soft gluon effects. Each choice of color structure
has, as a result, its own exponentiation for soft gluons \cite{BottsSt}.
Then, to next-to-leading-logarithm (NLL) 
it is possible to pick a color basis 
in which moments with respect to $z$ exponentiate, 
\begin{eqnarray}
\tilde{\Omega}_{ab}^{(IJ)}(n,y,\theta^*,Q^2)&=&\int_0^1 dz z^{n-1} 
\Omega_{ab}^{(IJ)}(z,y,\theta^*,\alpha_s(Q^2))
\nonumber \\ 
&=&H_{ab}^{(IJ)}(y,\theta^*,Q^2)
e^{E_{IJ}(n,\theta^*,Q^2)}\, ,
\label{omegaofn}
\end{eqnarray}
where the color-dependent exponents are given by
\begin{eqnarray}
E_{IJ}(n,\theta^*,Q^2)&=&-\int_0^1\frac{dz}{1-z}(z^{n-1}-1)
 \biggl [  \int_0^z {dy\over 1-y}\, g_1^{(ab)}[\alpha_s((1-y)(1-z)Q^2)]
\nonumber \\ &&
      +\, g_2^{(ab)}[\alpha_s((1-z) Q^2)]
+g_3^{(I)}[\alpha_s((1-z)^2 Q^2),\theta^*] \nonumber \\
&\ & \hbox{\hskip 1.0 true in}
+g_3^{(J)*}[\alpha_s((1-z)^2 Q^2),\theta^*]\, \biggr ]\, .
\label{Eofn}
\end{eqnarray}
The $g_i$ are finite functions of their arguments, and
the $H_{ab}^{(IJ)}$ are infrared safe expansions in $\alpha_s(Q^2)$. 
$g_1^{(ab)}$ and $g_2^{(ab)}$ are universal 
among hard cross sections and color
structures for given incoming partons $a$ and $b$, but depend on whether these partons
are quarks or gluons.
On the other hand, $g_3^{(I)}$ summarizes
soft logarithms that depend directly on color exchange in the hard scattering,
and hence also on the identities 
and relative directions  of the colliding partons (through $\theta^*$), both incoming and 
outgoing.  

Just as in
the case of Drell-Yan, to reach the accuracy of NLL in the exponents, 
we need $g_1$ only to two loops, with leading logarithms coming entirely from
its one-loop approximation,
and the functions $g_2$ and $g_3^{(I)}$ only to
a single loop.    More explicitly,
in the DIS scheme for incoming quarks \cite{CSt,CT2}
\begin{equation}
g_1^{(q {\bar q})} = 2 C_F \left ( {\alpha_s\over \pi} 
+ \frac{1}{2} K\; \left({\alpha_s\over \pi}\right)^2\right )\, ,
\label{g1def}
\end{equation}
with $K$ given by
\begin{equation}
K= C_A\; \left ( {67\over 18}-{\pi^2\over 6 }\right ) - {5\over 9}n_f\, ,
\label{Kdef}
\end{equation}
where $n_f$ is the number of quark flavors.  
 $g_2$ is given for quarks by
\begin{equation}
g_2^{(q {\bar q})}=-{3\over 2}C_F\; {\alpha_s\over\pi}\, .
\end{equation}
As pointed out in \cite{CT2}, one-loop contributions to
$g_3$ may always be absorbed into the one-loop contribution to $g_2$
and the two-loop contribution to $g_1$. Because
$g_3^{(I)}$ depends upon $I$, however, it is advantageous to keep
this nonfactoring process-dependence separate.  We shall
describe how it is determined below.

First, let us sketch how these results come about \cite{CLS}.
After the normal factorization of parton distributions, soft gluons cancel
in inclusive hard scattering cross sections.  When restrictions are
placed on soft gluon emission, however, finite logarithmic
enhancements remain, and it is useful to 
separate soft partons from the hard scattering (which is then constrained to
be fully virtual).  
Soft gluons  may be factored from the hard scattering
into a set of Wilson lines, or ordered exponentials,
from which collinear singularities in the initial state are eliminated, either
by explicit subtractions or by a suitable choice of gauge \cite{fact}.
Assuming that the lowest-order process is two-to-two,
there will be two incoming {\em and} two outgoing Wilson lines.\footnote{In
Drell-Yan and other electroweak annihilation processes, there is a
pair of incoming lines only.} The result, illustrated in Fig.\ 1b,
is of the form, $H_{ab}^{IJ}S_{IJ}$, summed over the same color basis as in
eq.\ (\ref{Qyfact}) above.

The resulting
hard scattering and soft-gluon functions both require renormalization,
which is organized by going to a basis in the space of color 
exchanges between the Wilson lines.  
The renormalization is carried out
by a counterterm matrix in this space of color tensors. 
For incoming and outgoing lines of equal masses, such analyses have
been carried out to one loop in \cite{BottsSt}, \cite{SotSt} and \cite{GK}
and to two loops in a related process in \cite{KK}.
For an underlying partonic process $a+b\rightarrow c+d$, 
we then construct an anomalous
dimension matrix $\Gamma^{(ab\rightarrow cd)}_{IJ}$, where the indices
$I$ and $J$ vary over the various color exchanges possible in the
partonic process.  The soft function  $S_{IJ}$ then satisfies the
renormalization group equation \cite{BottsSt}
\begin{equation}
\bigg( \mu {\partial \over \partial \mu}+\beta(g){\partial \over \partial g}\bigg )\,
S_{IJ}=-\bigg[\Gamma_{II'}\delta_{JJ'}+\delta_{II'}\Gamma_{JJ'}\bigg ]
S_{I'J'}\, .
\label{gammaRG}
\end{equation}
This resummation of soft logarithms is analogous to
singlet evolution in deeply inelastic scattering, which involves the mixing of operators,
and hence of parton distributions.  The general solution, even in moment
space, is given in terms of ordered exponentials which, however, may
be diagonalized at leading logarithm.
For the resummation of soft logarithms in 
QCD cross sections, the same general pattern
holds, with mixing between hard color tensors.  Leading 
soft logarithms,
however, are next-to-leading overall in moment space, which 
allows the exponentiation (\ref{omegaofn}) at this level.

Of course, the analysis is simplest for 
external quarks, and most complicated for external gluons.  It is
also possible to imagine a similar analysis when there are more than
two partons in the final state.  This would be necessary
if we were to treat threshold corrections to ${\bar{\rm p}}{\rm p}
\rightarrow Q{\bar Q}+ {\rm jet}$, for instance, but we have not attempted 
to explore such processes in detail.

Given a choice of incoming and outgoing partons, 
next-to-leading logarithms
in the moment variable $n$ exponentiate as in (\ref{omegaofn}) in the color tensor
basis that diagonalizes $\Gamma^{(ab\rightarrow cd)}_{IJ}$,
with eigenfunctions $\lambda_I$.  The
resulting soft function $g_3^{(I)}$ is then simply
\begin{equation}
g_3^{(I)}[\alpha_s((1-z)^2 Q^2),\theta^*]=-\lambda_I[\alpha_s((1-z)^2 Q^2),\theta^*]\, ,
\end{equation}
where the eigenfunctions are complex in general, and depend
on the directions of the incoming and outgoing partons, as shown.

\mysection{Applications to $q \bar{q} \rightarrow Q \bar{Q}$}

These considerations may be illustrated by heavy quark
production through light quark annihilation,
\begin{equation}
q(p_a)+{\bar q}(p_b) \rightarrow {\bar Q}(p_1) + Q(p_2)\, .
\end{equation}
Following Ref.\ \cite{mengetal}, we define invariants 
\begin{equation}
t_1=(p_a-p_1)^2-m^2, \quad u_1=(p_b-p_1)^2-m^2, \quad s=(p_a+p_b)^2\, ,
\end{equation}
with $m$ the heavy quark mass, which satisfy
\begin{equation}
s+t_1+u_1=0
\end{equation}
at partonic threshold. 
In this case,
as in elastic scattering \cite{BottsSt,KK}, the anomalous
dimension matrix is only two-dimensional.  

In a color tensor basis
of singlet and octet exchange in the $s$ channel, the anomalous dimension
of eq.\ (\ref{gammaRG}) is,
\begin{eqnarray}
\Gamma_{11}&=&-\frac{\alpha_s}{\pi}C_F(L_{\beta}+1+\pi i),
\nonumber \\
\Gamma_{21}&=&\frac{2\alpha_s}{\pi}
\ln\left(\frac{u_1}{t_1}\right),
\nonumber \\ 
\Gamma_{12}&=&\frac{\alpha_s}{\pi}
\frac{C_F}{C_A} \ln\left(\frac{u_1}{t_1}\right),
\nonumber \\
\Gamma_{22}&=&\frac{\alpha_s}{\pi}\left\{C_F
\left[4\ln\left(\frac{u_1}{t_1}\right)-L_{\beta}-1-\pi i\right]\right.
\nonumber \\ &&
\left.+\frac{C_A}{2}\left[-3\ln\left(\frac{u_1}{t_1}\right)
-\ln\left(\frac{m^2s}{t_1u_1}\right)+L_{\beta}+\pi i \right]\right\}\, ,
\label{gammaoneloop}
\end{eqnarray}
where subscript 1 denotes singlet,
and 2 octet.   $L_\beta$ is the familiar velocity-dependent
eikonal function
\begin{equation}
L_{\beta}=\frac{1-2m^2/s}{\beta}\left(\ln\frac{1-\beta}{1+\beta}
+i \pi \right)\, ,
\end{equation}
with $\beta=\sqrt{1-4m^2/s}$.
The matrix depends, as expected, on the directions of the Wilson
lines, which may be reexpressed in terms of ratios of
kinematic invariants for the partonic scattering.  For arbitrary
$\beta$ and fixed scattering angle, we must solve for the relevant 
diagonal basis 
of color
structure, and determine the eigenvalues.   
However, $\Gamma$ is already diagonalized in the
$s$-channel octet-singlet basis 
at ``absolute" threshold, $\beta=0$,
and, more generally, when the parton-parton c.m. scattering angle is
$\theta^*=90^\circ$ (where $u_1=t_1$). 

It is of interest, of course, to compare the one-loop expansion of
our results to known one-loop calculations, at the level of 
next-to-leading order.
We may give our result as a function of $z$, since the inverse
transforms are trivial.
They are found in terms of the Born cross section,
the one-loop factoring contributions of $g_1^{(q{\bar q})}$ 
and $g_2^{(q{\bar q})}$,
and $\Gamma_{22}$.
In the DIS scheme the result is
\begin{eqnarray}
\sum_{IJ}\; {\Omega^{(IJ)}_{q{\bar q}}(z,u_1,t_1,s)}{^{(1)}}
&=&
\sigma_{\rm Born}
\frac{\alpha_s}{\pi} \frac{1}{1-z}\left\{C_F\left[2\ln(1-z)
 +\frac{3}{2} \right.\right.\nonumber \\
&\ & \quad \quad 
\left.\left.+8\ln\left(\frac{u_1}{t_1}\right) 
-2-2 L_{\beta}+2\ln\left(\frac{s}{\mu^2}\right)\right. \right]
\nonumber \\
&\ & \quad 
\left.+C_A\left[-3\ln\left(\frac{u_1}{t_1}\right)+L_{\beta}
-\ln\left(\frac{m^2s}{t_1 u_1}\right)\right]\right\}\, .
\nonumber\\
\label{longeq}
\end{eqnarray}
Here $\mu$ is the factorization scale, and the logarithm of $s/\mu^2$
describes the evolution of the parton distributions.
This result cannot be compared directly to
the exact one-loop results of \cite{heavycalcs} for arbitrary $\beta$,   
where the singular behavior is given in terms of the variable $s_4$,
with
\begin{equation}
s_4= (p_2+k)^2-m^2 \approx 2p_2\cdot k\, ,
\end{equation}
where $k=p_a+p_b-p_1-p_2$ is the momentum carried away by the gluon.  At
partonic threshold, both $s_4$ and $(1-z)$ vanish, but even for small $s_4$, angular
integrals over the gluon momentum with $s_4$ held fixed
are rather different than those with
$1-z \approx 2(p_1+p_2)\cdot k/s$ held fixed.  

Nevertheless, the cross sections become identical in the 
$\beta\rightarrow 0$ limit, where we may make a direct
comparison.
Near $s=4m^2$, we  may identify $2m^2(1-z)=s_4$, and 
(2.6) reduces to the $\beta\rightarrow 0$
limit of eq.\ (30) of \cite{mengetal}.  It is also worth
noting that even for $\beta>0$, the two cross
sections remain remarkably close, differening only
at first nonleading logarithm in the abelian ($C_F^2$)
term,
due to the interplay of angular integrals 
with leading singularities for the differing treatments of
phase space.  

As for the Drell-Yan cross section, our analysis 
applies not only to absolute threshold for the production of the
heavy quarks ($\beta=0$), but also to partonic threshold
for the production of moving heavy quarks.
When $\beta$ nears unity, however,
the anomalous dimensions themselves develop (collinear) singularities
associated with the fragmentation of the heavy quarks, which
in principle may be factored into nonperturbative
fragmentation functions.  

In summary, we have illustrated the application of a general method for
resumming next-to-leading logarithms at partonic threshold
in QCD cross sections.  Possible
extensions include, of course, heavy-quark production through
gluon fusion, and dijet production.  Extensions to multijet production
are also possible.  We reserve estimates of the phenomenological
importance of these nonleading terms to future work, but we
hope that whether they give small contributions or large,
the method will improve the reliability of perturbative QCD calculations
for hard scattering cross sections.

\medskip

\noindent
{\em Acknowledgements}.  We wish to thank Jack Smith and
Eric Laenen for useful conversations.
This work was supported in part by the National Science Foundation under
grant PHY9309888.

\newpage

\bigskip
\noindent{\Large \bf Figure Caption}
\medskip

Cut diagram illustrating momentum configurations that give rise to
threshold enhancements in heavy quark production.  (a) General
factorization theorem.  Away from partonic threshold all singularities
in the ``short-distance" subdiagram $H/S$ cancel;
(b) Expanded view of $H/S$ near threshold, showing the factorization
of soft gluons onto eikonal (Wilson) lines from incoming and outgoing
partons in the hard subprocess.  $H_I$ and $H^*_J$ represent the
remaining, truly short-distance, hard scattering.


\begin{thebibliography}{99}

\bibitem{oldDY} G.\ Sterman, Nucl.\ Phys.\ B281 (1987) 310;
 S.\ Catani and L.\ Trentadue, 
Nucl.\ Phys.\ B327 (1989) 323.

\bibitem{heavyquarkresum}  E.\ Laenen, J.\ Smith and W.L.\ van Neerven, 
Nucl.\ Phys.\  B369 (1992) 543;
Phys.\ Lett.\ B321 (1994) 254;
E.L.\ Berger and H.\ Contopanagos, Phys.\ Lett.\  B361 (1995) 115;
 Argonne preprint ANL-HEP-95-82,
hep-ph/9603326;
N.\ Kidonakis and J.\ Smith, Phys.\ Rev.\ D51 (1995) 6092; 
S.\ Catani, M.L.\ Mangano, P.\ Nason and L.\ Trentadue,
CERN preprint CERN-TH/96-21, hep-ph/9602208.


\bibitem{heavycalcs} P.\ Nason, S.\ Dawson and R.K.\ Ellis,
Nucl.\ Phys.\ B303 (1988) 607; W.\ Beenakker, H.\ Kuijf, W.L.\ van Neerven,
and J.\ Smith Phys.\ Rev.\ D40 (1989) 54; W.\ Beenakker, W.L.\ van
Neerven, R.\ Meng, G.A.\ Schuler, and J.\ Smith, Nucl.\ Phys.\ 
B351 (1991) 507.

\bibitem{jetcalcs}  F.\ Aversa, P.\ Chiappetta, M.\ Greco
and J.-Ph.\ Guillet, Phys.\ Lett.\ B211 (1988) 465;
Phys.\ Rev.\ Lett.\ 65 (1990) 401; Z.\ Phys.\ C46 (1990) 253;
F.\ Aversa, P.\ Chiappetta, L.\ Gonzales, M.\ Greco
and J.-Ph.\ Guillet,
Z.\ Phys.\ C49 (1991) 459;
S.D.\ Ellis, Z.\ Kunszt and D.E.\ Soper,
Phys.\ Rev.\ D40 (1989) 2188; Phys.\ Rev.\ Lett.\  64 (1990) 2121;
{\it ibid} 69 (1992) 1496;  Z.\ Kunszt and D.E.\ Soper, Phys.\ Rev.\
D46 (1992) 192.

\bibitem{fact} J.C.\ Collins, D.E.\ Soper and G.\ Sterman, in
{\it Perturbative Quantum Chromodynamics},
ed.\ A.H.\ Mueller (World Scientific, Singapore, 1989) p.\ 1.

\bibitem{expon}  J.G.M.\ Gatheral, Phys.\ Lett.\ 133B (1983) 90;
J.\ Frenkel and J.C.\ Taylor, Nucl.\ Phys.\ B246 (1984) 231;
G.P.\ Korchemsky and A.V.\ Radyushkin, Phys.\ Lett.\ B171 (1986) 459;
J.C. Collins, in {\it Perturbative Quantum Chromodynamics},
ed.\ A.H.\ Mueller (World Scientific, Singapore, 1989), p.\ 573.

\bibitem{CLS} H. Contopanagos, E.\ Laenen, and G. Sterman, in preparation.

\bibitem{LaenenSterman} E.\ Laenen and G.\ Sterman, in 
proceedings of {\it The Fermilab
Meeting, DPF 92}, 7th  meeting of the 
American Physical Society Division of Particles
and Fields (Batavia, IL, 1992), ed.\
C.H.\ Albright {\it et al.} (World Scientific,
Singapore, 1993), p.\ 987.

\bibitem{BottsSt} J.\ Botts and G.\ Sterman, Nucl.\ Phys.\ B325 (1989) 62.

\bibitem{CSt} H.\ Contopanagos and G.\ Sterman, Nucl.\ Phys.\ B400 (1993) 211; 
B419 (1994) 77.

\bibitem{CT2}  S.\ Catani and L.\ Trentadue, B353 (1991) 183.

\bibitem{SotSt} M.\ Sotiropoulos and G.\
Sterman,  Nucl.\ Phys.\ B419 (1994) 59.

\bibitem{GK} G.P. \ Korchemsky, Phys.\ Lett.\ B325 (1994) 459.

\bibitem{KK}  I.A.\ Korchemskaya and G.P.\ Korchemsky,
Nucl.\ Phys.\ B437 (1995) 127. 

\bibitem{mengetal} R.\ Meng, G.A.\ Schuler, J.\ Smith and W.L.\ van Neerven,
Nucl.\ Phys.\  B339 (1990) 325.

\end{thebibliography}
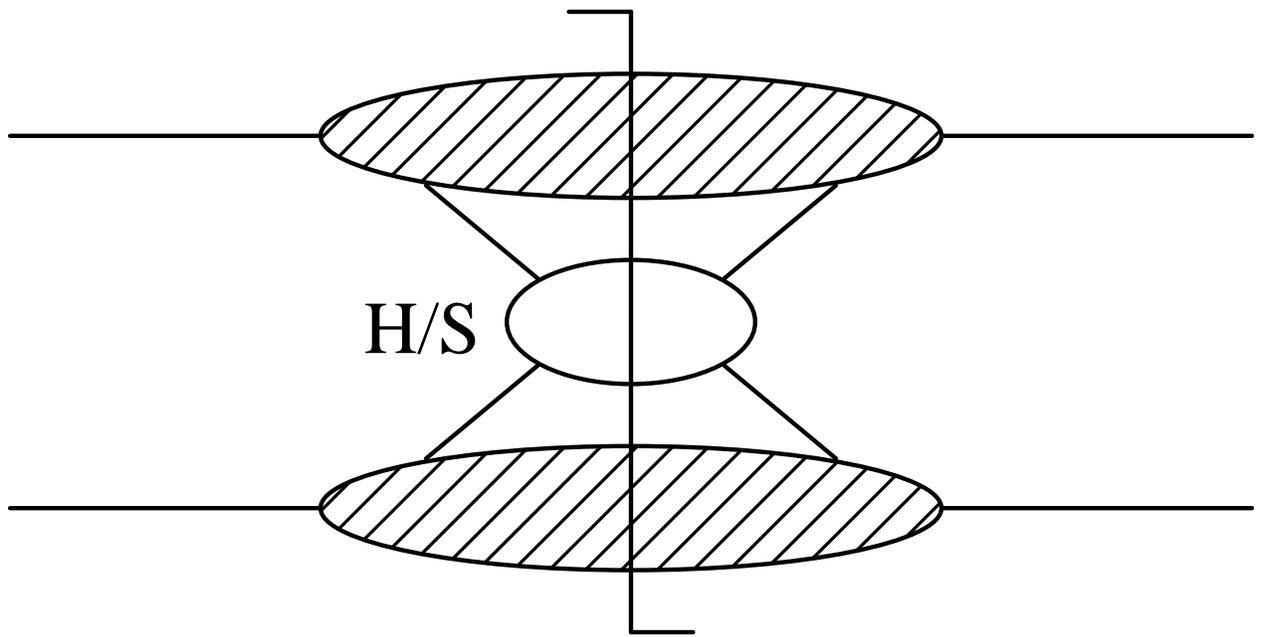
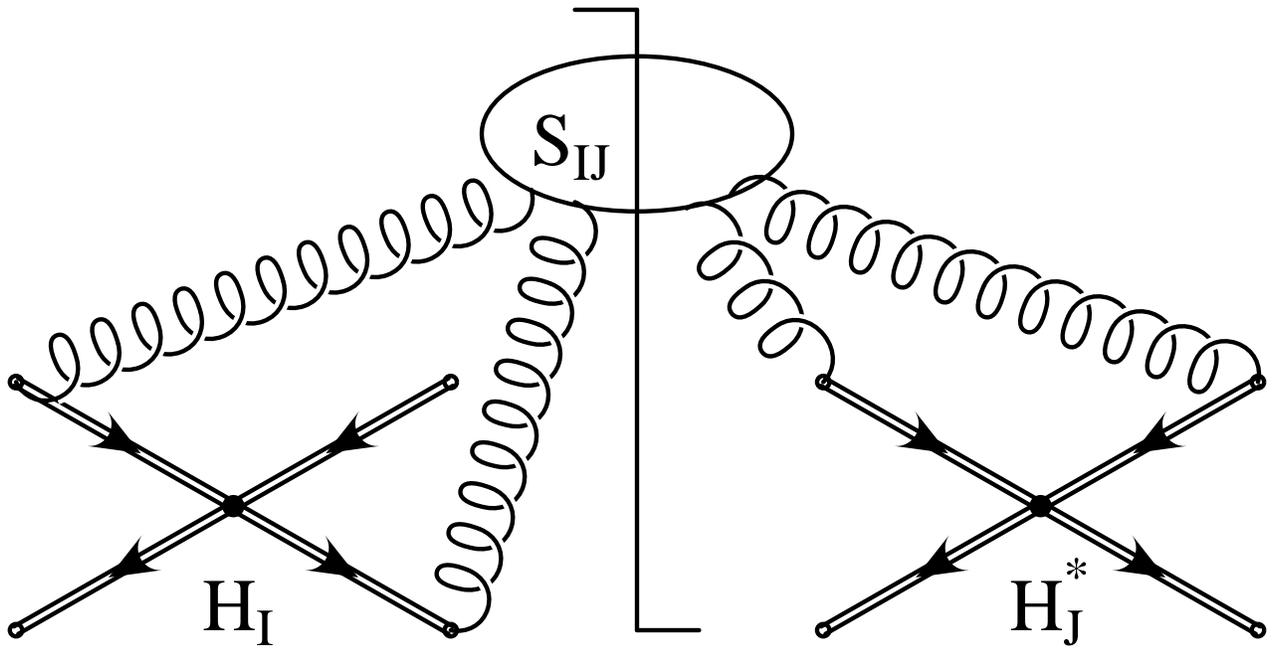
\end{document}